\documentclass[journal]{IEEEtran}
\usepackage{epsfig,mathtools,amsmath,amssymb}
\usepackage{color}
\usepackage{dirtytalk}
\usepackage{url}
\usepackage{mathtools}
\usepackage{subfigure}
\usepackage{multirow}
\usepackage{verbatim}
\usepackage{algorithm, algorithmic}
\usepackage{booktabs}
\usepackage{bbm}

\usepackage{hyperref}
\definecolor{mydarkblue}{rgb}{0,0.08,0.45}
\hypersetup{colorlinks,linkcolor=red,urlcolor=mydarkblue,linktoc=page}

\newcommand{\params}{\boldsymbol{\theta}}

\definecolor{gray}{rgb}{0.5,0.5,0.5}
\newcommand{\worse}[1]{\textcolor{gray}{#1}}

\newcommand{\revised}[1]{\textcolor{black}{#1}}
\ifCLASSINFOpdf
\else
\fi
\begin{document}
\title{Optimizing Tandem Speaker Verification and Anti-Spoofing Systems}

\author{Anssi~Kanervisto,
        Ville~Hautam\"aki,~\IEEEmembership{Member,~IEEE,}
        Tomi~Kinnunen
        and~Junichi Yamagishi,~\IEEEmembership{Senior Member,~IEEE}% <-this % stops a space
\thanks{A. Kanervisto, V. Hautam\"aki and T. Kinnunen are with School of Computing, University of Eastern Finland, Joensuu, Finland.}% <-this % stops a space
\thanks{V. Hautam\"aki is also with the National University of Singapore.}% <-this % stops a space
\thanks{J. Yamagishi is with National Institute of Informatics, Tokyo, Japan.}
\thanks{Corresponding author: Anssi Kanervisto (\texttt{anssk@uef.fi}).}
}


\markboth{Optimizing tandem speaker verification and anti-spoofing systems}%
{Optimizing tandem speaker verification and anti-spoofing systems}



\maketitle

\begin{abstract}
As automatic speaker verification (ASV) systems are vulnerable to spoofing attacks, they are typically used in conjunction with spoofing countermeasure (CM) systems to improve security. For example, the CM can first determine whether the input is human speech, then the ASV can determine whether this speech matches the speaker’s identity. The performance of such a tandem system can be measured with a  tandem detection cost function (t-DCF). However, ASV and CM systems are usually trained separately, using different metrics and data, which does not optimize their combined performance. In this work, we propose to optimize the tandem system directly by creating a differentiable version of t-DCF and employing techniques from reinforcement learning. The results indicate that these approaches offer better outcomes than finetuning, with our method providing a 20\% relative improvement in the t-DCF in the ASVSpoof19 dataset in a constrained setting.

%The spoofing countermeasure (CM) systems in automatic speaker verification (ASV) are not typically used in isolation of each other. These systems can be combined, for example, into a cascaded system where CM produces first a decision whether the input is synthetic or bonafide speech. In case the CM decides it is a bonafide sample, then the ASV system will consider it for speaker verification. End users of the system are not interested in the performance of the individual sub-modules, but instead are interested in the performance of the combined system. Such combination can be evaluated with tandem detection cost function (t-DCF) measure, yet the individual components are trained separately from each other using their own performance metrics. In this work we study training the ASV and CM components together for a better t-DCF measure by using reinforcement learning. We demonstrate that such training procedure indeed is able to improve the performance of the combined system, and does so with more reliable results than with the standard supervised learning techniques we compare against.
\end{abstract}


% \fi
\IEEEpeerreviewmaketitle

\section{Introduction}
    An \textit{automatic speaker verification} (ASV) system attempts to verify if a given speech utterance matches the claimed identity~\cite{Reynolds95_speaker}. As with any machine learning system, this system is susceptible to malicious inputs such as replay attacks~\cite{kinnunen2017asvspoof} or speech synthesis~\cite{todisco2019asvspoof}, where the attacker aims to fool the ASV system with crafted audio samples. \textit{Spoofing countermeasure} (CM) systems aim to detect these crafted audio samples, and thus improve security when combined with an ASV system~\cite{todisco2019asvspoof}. This improvement is achieved by separately training the two systems, then using them in conjunction with each other and evaluating their performance using a \textit{tandem detection cost function} (t-DCF)~\cite{kinnunen2020tandem}.
    
    %\todo{rewrite} \textit{Automatic speaker verification} (ASV) systems attempt to verify if a given speech utterance matches the claimed identity \cite{Reynolds95_speaker}. Alone such systems are susceptible to malicious attacks via replay attack \cite{kinnunen2017asvspoof} or speech synthesis \cite{todisco2019asvspoof}, where an attacker attempts to fool the ASV system with crafted audio samples. To combat this, \textit{spoofing countermeasure} (CM) systems aim to detect these malicious samples from genuine, \say{bonafide} samples. These systems can then be used along with ASV systems to detect and discard access attempts with \textit{e.g.} synthetic speech, to improve the security of the system \cite{todisco2019asvspoof}.
    
    %\todo{rewrite}The ASV and CM systems are trained in isolation from each other, and later combined by running test trials through both systems. To measure performance of such a combined system, the authors of \cite{kinnunen2018t} proposed \textit{tandem detection cost function} (t-DCF) to evaluate the performance of the combined system performance. This performance metric takes into account different error situations that arise from having two different systems and assigns different costs for these errors. It is a generalization of the widely adopted \textit{detection cost function} (DCF) \cite{brummer2006application} involving an extra, \say{spoof} class and two (rather than one) systems being evaluated.
    
    Although they are evaluated using this tandem metric, the original ASV and CM systems are not trained to minimize the t-DCF. They are instead trained using their respective losses in their individual tasks, with no consideration given to the existence of the other system that it will be combined with. While there are no immediate downsides to this practice (these tandem systems have achieved high accuracy~\cite{todisco2019asvspoof}), there could be room for improvement. Some attack systems used to generate spoof samples could fool the CM but may be easily detected by the ASV system, as is the case with system A17 in the ASVspoof19 dataset~\cite{wang2020asvspoof}. 
    
    Alas, as the t-DCF metric requires hard counts of misses and false accepts, it becomes a non-differentiable metric that cannot be directly optimized with gradient descent methods. Fortunately, metrics can be \say{softened} (made differentiable) by approximating the non-differentiable operations with differentiable ones, such as \textit{maximal figure-of-merit} \cite{kukanov2016deep, gao2006maximal} or \textit{hinge-loss}~\cite{sizov2017direct, bishop2006pattern}. Here we use the same technique and extend a soft (differentiable) version of the \emph{detection cost function} (DCF) metric~\cite{mingote2019optimization} to accommodate the t-DCF. However, using approximations in place of exact counts leads to inaccuracies, and these soft metrics may require tuning for efficient learning~\cite{gao2006maximal}.
    
    % The original ASV and CM systems are not trained to minimize the t-DCF metric: they are trained to minimize the errors in their respective tasks, completely oblivious of the existence of the other systems and the final use-case. Our aim is to improve the metric of the final task, namely the t-DCF. Unfortunately, to compute t-DCF, one has to compute error-rates with hard decisions, which is a non-differentiable operation, preventing updating model parameters with gradient descent. One approach for optimizing count-based metrics relies on \say{softening} the counts with soft (differentiable) versions, \textit{e.g.} by methods based on \textit{maximal figure-of-merit} \cite{kukanov2016deep, gao2006maximal} or using \textit{hinge-loss} to approximate error counts \cite{sizov2017direct, bishop2006pattern}. While shown to be effective optimization techniques, these methods involve functions that tend to require tuning of parameters for efficient learning \cite{gao2006maximal}.
    
    While \textit{reinforcement learning} (RL)~\cite{sutton2018reinforcement} is often associated with games and robotics, it has also, perhaps surprisingly, been used in conjunction with multi-step supervised tasks such as \emph{sequence generation}~\cite{yu2017seqgan}. In this study, 
    we consider an entire novel application area of RL to \emph{optimize} tandem systems---here, ASV, CM, and their combination. Even if RL 
    methods have been applied for this purpose to small datasets~\cite{jang2017categorical} and adversarial attacks~\cite{gupta2020reinforcement}, the application of RL to classifier optimization remains largely unexplored. Compared to differentiable objective functions that are typically optimized using gradient-based methods, RL methods (in particular, REINFORCE~\cite{williams1992simple}) are widely applicable for optimizing \emph{any} objective that involves sampling from a probability distribution. REINFORCE offers a zero-bias estimate of the gradient for minimizing the objective.

    Because REINFORCE can be directly applied to non-differentiable objective functions without approximations, it may lead to different results compared to optimizing soft versions of the objectives. To this end, we propose optimizing tandem ASV and CM systems with the REINFORCE algorithm. We pre-train the two systems and then optimize them in tandem using REINFORCE, a softened evaluation metric, and, as a baseline result, cross-entropy loss for the same data. The key advance of this work is the process of optimizing the entire tandem system at once; moreover, we are first to optimize t-DCF directly. In contrast to preliminary results reported in~\cite{kanervisto2020initial}, here we conduct experiments using state-of-the-art ASV and CM systems, perform more theoretically sound REINFORCE training using score calibration, and include a soft version of the t-DCF metric in the comparison.
    
    \revised{Our contributions can be summarized as follows:
    \begin{enumerate}
        \item We propose jointly training a tandem ASV and CM system using methods from RL literature.
        \item We present a novel reward function for optimizing t-DCFs using RL and discuss the importance of score calibration for RL training.
        \item We analyze the proposed method against baseline solutions with the ASVSpoof19 dataset; these baselines include a soft (differentiable) version of t-DCF, which we derive based on previous work on soft decision cost functions.
        \item We study the results of each spoofing attack and the generalization of the different joint training methods.
        \item We study how the cross-task performance changed during tandem training (i.e., did ASV learn to detect spoof samples during tandem training?).
    \end{enumerate}}
    
    %We propose to optimize tandem system with ASV and CM systems to perform secure speaker verification using the REINFORCE algorithm. We pre-train disjoint ASV and CM systems which are then fine-tuned for the spoofing-robust speaker verification task with REINFORCE using rewards derived from the t-DCF cost parameters. We then compare these against standard supervised approaches for fine-tuning such systems. Our novelty is two-fold. First, the use of REINFORCE on a system of multiple individual parts (ASV and CM) allows us to analyze how individual components change during training. Second, this is the first attempt of optimizing t-DCF directly for spoofing-robust speaker verification system.

\section{Tandem Detection Cost Functions (t-DCFs)}
\label{sec:tdcf}
    %\todo{UNDER WORK WITH TOMI, DON'T READ/COMMENT YET}
    
    The t-DCF~\cite{kinnunen2020tandem} extends the conventional DCF \cite{greenberg2020-two-decades} used in speaker recognition studies. We briefly review each approach. 
    
    \subsection{Conventional Detection Cost Function (DCF)---one system, two classes}
        A \emph{binary classifier} (or a \emph{detector}) is any function $g: \mathcal{X} \rightarrow \{0,1\}$ that assigns a predicted class label, $\hat y = g(x)$, to a given input $x \in \mathcal{X}$. One can view $g(\cdot)$ as a device that either accepts ($\hat y=1$) or rejects ($\hat y=0$) a given hypothesis. Since there are only two possible outcomes (classes), accepting one hypothesis means rejecting its complement and vice versa. In practice, one first computes a real-valued score (soft decision), $s \in \mathbb{R}$, and compares it to a pre-set threshold, $\tau$, to reach the binary decision. The hypothesis is accepted if and only if $s > \tau$. In ASV \cite{reynolds95-speaker-rec}, $x$ comprises a pair of enrollment and test utterances and the hypothesis is that the same speaker is present in the two. In voice anti-spoofing, $x$ is a single utterance and $g(\cdot)$ tests whether or not $x$ is true human speech or a spoofing attack. %In the case of \emph{voice anti-spoofing}, $X$ is a single utterance and the null hypothesis is that $X$ originates from a bonafide (human) user. The classifier acts as a \emph{countermeasure} (CM) to detect spoofing attacks.
    
        Since neither the classifier nor its input $x$ are perfect, the prediction $\hat y$ may differ from the actual (true) class, $y \in \{0,1\}$. There are two possible errors. A \emph{false accept} happens when $\hat y=1$ but $y=0$, and a \emph{miss} (false rejection) happens when $\hat y=0$ but $y=1$. By viewing $y$ and $\hat y$ as instances of random variables $Y$ and $\hat Y$, the miss and false accept \emph{rates} can be are defined as the conditional probabilities $P_\text{miss} \equiv p(\hat Y=0|Y=1)$ and $P_\text{fa} \equiv p(\hat Y=1|Y=0)$, respectively. Their empirical estimates are obtained by simply counting the errors (i.e., counting the ratio of errors with a threshold $\tau$).
        
        %When viewed as a decision problem, one assigns \emph{detection costs} to each one of these two error types, along with with \emph{a priori} probability of each class. 
        %An automatic speaker verification system compares enrollment and test utterances in terms of their speaker similarity and takes a hard \texttt{ACCEPT} (same speaker) or \texttt{REJECT} (different speaker) action. As a binary any classifier, ASV systems are prone to two different types of errors, \emph{false alarms} (false accepts) and \emph{misses} (false rejections). 
        The miss and false accept rates can be considered proxies for user convenience and security, respectively. Neither should be compromised, but the trade-offs between the two can be adjusted (by selecting different values of $\tau$), and there might be different preferences for their relative severity depending on the application. Performance is therefore gauged using a flexible, parameterized DCF \cite{greenberg2020-two-decades,brummer2006application},
            \begin{equation}
                \begin{split}
                \label{eq:basic-dcf}
                \text{DCF} & = \rho_\text{tar}C^\text{asv}_\text{miss}P_\text{miss}^\text{asv} + \rho_\text{non}C_\text{fa}P_\text{fa}^\text{asv}\\
                & = \rho_\text{tar}C^\text{asv}_\text{miss}P_\text{miss}^\text{asv} + (1-\rho_\text{tar})C_\text{fa}P_\text{fa}^\text{asv},\\
                \end{split}
            \end{equation}
        where $\rho_\text{tar} \equiv p(Y=1)$ and $\rho_\text{non} \equiv p(Y=0)=1-\rho_\text{tar}$ are the \emph{prior probabilities} of positive (target) and negative (non-target) class, while $C_\text{miss}, C_\text{fa} > 0$ are \emph{costs} of a miss and a false accept, respectively. We introduced the superscript `asv' to emphasize that these errors are those of an ASV system (generally, any single classifier). The DCF parameters, $(C_\text{miss}, C_\text{fa}, \rho_\text{tar})$ are set in advance and remain fixed in a given evaluation setup. Note that the prior $\rho_\text{tar}$ reflects prior uncertainty concerning the class label: the prior can (and usually is) different from the empirical proportion of test trials in an evaluation dataset.
    
    \subsection{t-DCF---two systems, three classes}
    
        While \eqref{eq:basic-dcf} is well-suited to assessing the performance of regular ASV systems in which only target (same speaker) and non-target (different speaker) trials are encountered, it does not consider the impact of spoofing attacks. The t-DCF framework \cite{kinnunen2018t,kinnunen2020tandem} expands the scope by adding a third class (spoofing attack) and another classifier, a spoofing CM. The CM aims to discriminate between true utterances and spoofing attacks. To leverage the benefits of each system, we place a CM in front of an ASV system as a bonafide/spoof gate. The t-DCF metric assesses the performance of this cascaded, or tandem, system.
        
        While conventional ASV systems have two possible errors, the tandem system may now experience four different types of errors: (a) a target user is accepted by the CM but rejected by the ASV system, (b) a non-target user gets accepted by both systems, (c) a spoofing attack gets accepted by both systems, or (d) a target user is rejected by the CM. With this motivation, \cite{kinnunen2020tandem} extended the DCF to a t-DCF, where
                \begin{equation}
                \begin{split}
                \label{eq:new_tdcf}
                \text{t-DCF} &= C_\text{miss} \cdot \rho_\text{tar} \cdot (P_\text{a} + P_\text{d}) \\
                    &+ C_\text{fa} \cdot \rho_\text{non} \cdot P_\text{b} \\
                    &+ C_\text{fa,spoof} \cdot \rho_\text{spoof} \cdot P_\text{c},
                \end{split}
            \end{equation}
        where $P_\bullet$ denotes the error rates of the four cases noted above (obtained by treating the CM and ASV decisions independently) and $\rho_\text{spoof}$ is the prior probability of a spoofing attack. Since there are now three mutually exclusive classes, $\rho_\text{tar}+\rho_\text{non} + \rho_\text{spoof}=1$, the three cost parameters, $C_\text{miss}$, $C_\text{fa}$, and $C_\text{fa,spoof}$, are now the detection costs of the tandem system. The first two have the same meaning as their corresponding ASV-system costs in \eqref{eq:basic-dcf}, while $C_\text{fa,spoof}$ is the cost of falsely accepting a spoofing attack. The tandem cost used in this work is thus presented by five parameters $(C_\text{miss}, C_\text{fa}, C_\text{fa,spoof}, \rho_\text{tar}, \rho_\text{non})$.
        
        \revised{While the previous version of t-DCF~\cite{kinnunen2018t} contained four separate costs (two miss cases, two false-accept cases), this formulation was later revised~\cite{kinnunen2020tandem}. For completeness, \eqref{eq:new_tdcf} includes the miss rates of both systems, but the impact of a miss is ultimately the same from the end-user's perspective regardless of which subsystem caused the error. For this reason, the two miss terms share the same cost and prior terms. If the cost and priors of false acceptance terms are the same for both scenarios, the t-DCF becomes the original DCF \eqref{eq:basic-dcf}. However, the system designer may want to set the cost of falsely accepting a spoofing attack higher than the cost of a miss, as accepting an intentional attack (spoofing attack) is more undesirable than accepting a non-target speaker.}
        
        Each of the four error rates in \eqref{eq:new_tdcf} now depends on two different thresholds---one for the CM system and another for the ASV system. The authors of \cite{kinnunen2020tandem} discuss two different use cases: one where the ASV remains a fixed black box (fixed threshold and error rates that cannot be adjusted) and another where both the CM and the ASV systems are adjustable. The former, \emph{ASV-constrained} t-DCF, is suited for scenarios where the two classifiers are developed by different parties. The latter, \emph{unconstrained} t-DCF, is applicable when both classifiers are adjustable. \revised{A CM-constrained setup, where the CM is a fixed system, has not been considered in the context of t-DCF. The t-DCF was designed to evaluate the performance of CM systems in combination with ASV systems and to further develop CM systems; ASV systems are provided by a separate research community with rapid methodological progress~\cite{zeinali2019but, garciaromero2020magneto}. In a CM-constrained t-DCF setup, one would assume the CM system to be is fixed and the ASV would be adjusted to fit the CM, which is contrary to the assumption of ASV being a fixed black box.}

        % Removed as we do not use this normalization but rather the one in ASVspoof19 competition
        %The cost is further normalized so that a value of 1 corresponds to an uninformative default (or reference) system, obtained by dividing \eqref{eq:new_tdcf} by $\min \{C_\text{fa}\cdot \rho_\text{non} + C_\text{fa,spoof}\cdot \rho_\text{spoof},C_\text{miss}\cdot \rho_\text{tar}\}$. \todo{Is this same as in code? Check! Do we need this normalization? Remove } 
    
    \subsection{Optimizing DCF or t-DCF}
        Both \eqref{eq:basic-dcf} and \eqref{eq:new_tdcf} are helpful for performance \emph{evaluation} but not \emph{optimization}. All the error rates needed for computing these metrics are obtained by counting errors, leading to a generally non-differentiable optimization problem. We address this challenge using two approaches. First, we design a \textit{soft t-DCF} metric using approximations for the hard error counts \cite{kukanov2016deep, silver2016mastering, gao2006maximal}, which we can then differentiate and optimize for directly (Section \ref{sec:method2}). However, due to these approximations, the softened metric does not match the original t-DCF. Instead, we can optimize directly for the original t-DCF metric without approximations using methods from reinforcement learning~\cite{sutton2018reinforcement}. Before detailing our approaches (Sections \ref{sec:method} and \ref{sec:method2}), we provide a brief background on RL.
    
% \section{Optimizing for a specific threshold}
    % \todo{Tomi: You had some ideas regarding this and we have nudges in DETs toward this. Do you want to add a word or two about it here? Sounds like optimizing t-DCF would/should lead to optimizing that specific point on DET-curve}

\section{Reinforcement learning and policy gradients}
    \label{sec:rl}
    % Anssi: This section has been rewritten:
    % - Removed stuff that we do not use in this paper
    % - Clarify focus and use above ASV systems as an example from a start
    In this section, we cover the terminology and base concepts of RL and, more specifically, methods using the \textit{policy gradient theorem} (PG) \cite{sutton2000policy}, which lies at the core of our proposed approach. A key feature of PG is \textit{the concept of sampling actions}: instead of designing a system to output scores and then determining the decision threshold, RL trains systems that pick the actions themselves. These actions can be chosen deterministically or via stochastic sampling. This practice allows the RL system to naturally \textit{explore} different paths of actions. Once a promising action has been found, the system is updated to choose this action more frequently (\say{exploit} knowledge) \cite{sutton2018reinforcement}. Unlike back-propagation-based loss-minimization training, PG methods can be used to train systems that are non-differentiable, much like the t-DCF metric, by estimating the true gradient which can then be used to optimize for the objective~\cite{sutton2000policy}.

    \subsection{Reinforcement learning (RL)}
        RL aims to solve problems formulated as \textit{Markov decision processes} (MDPs), which consist of five elements. Using RL notation and the speaker verification setup as an example, the ASV system $\pi: \mathcal X \rightarrow \{\text{\texttt{ACCEPT}, \texttt{REJECT}}\}$  (\textbf{policy}) chooses an action $a \in \{\text{\texttt{ACCEPT}, \texttt{REJECT}}\}$\footnote{In this work, this is same as the prediction label $\hat y$, but we use notation general to RL.} given a state\footnote{Note that we overload the notion of $x \in \mathcal X$ being input. For this paper, the meaning of state and input is the same. The convention is to use $s$ to represent the state in RL literature, but we use notion $s$ to represent the score in this work.} $x \in \mathcal X$, either deterministically or stochastically. Here $\pi$ is usually a neural network with millions of parameters, and the state is a pair of utterances the ASV will compare. After choosing an action, the policy is rewarded with $r = R(x, a)$, where the function $R$ is determined by the designer, where the reward can be positive for the correct choice and zero otherwise, for example. The task for the policy is simple: pick the action with the highest expected reward. To optimize (minimize) an error metric, one can maximize its negative, which is the same as minimizing the original error.
        
        In the generic case, an MDP lasts for multiple \textbf{timesteps} $t \in \mathbb{N}$, starting from an initial state sampled from a distribution $x_0 \sim p(x_0)$ and, potentially, ending at a \textbf{terminal} state $x_T$. The subscripts $0$ and $T$ denote the timestep in which the state was encountered, and $T \in \mathbb{N}$ represents the final timestep when the \textbf{episode} ends. After each action, the MDP transitions to the next state according to the \textbf{transition dynamics} $x' \sim p(x' | x, a)$, where $x'$ denotes the state succeeding $x$. The goal in MDPs is to obtain an \textit{optimal policy} $\pi^*$ which maximizes the expected sum of rewards
        %
        %\begin{equation}
        %    \pi^* := \arg\max_\pi \mathbb E \left [ \sum_{t=0}^{T} r_t \right ] ~ .
        %\end{equation}
        %
        %\textcolor{blue}{(JY comment:this equation does not intuitively show relationships between variants and rewards. Can you try to clearly show it, please? One option is to show them using consistent subscripts like this. That also helps a gap between Eq. 3 and 8. 
        %
        %\begin{align}
        %   \pi^*   &\equiv \arg\max_\pi \mathbb E \left [ \sum_{t=0}^{T} r_t \right ]  \\
        %    \tilde{r}_t     &= R(\tilde{x}_t, \tilde{a}_t) \\
        %    \tilde{x}_t     &\sim p(x_t | x_{t-1}, a), \tilde{x}_0 \sim p(x_0)\\
        %    \tilde{a}_t     &\sim \pi(a_t | x_t) ~ .
        %\end{align}
        \begin{equation}
            \pi^*   \equiv \arg\max_\pi \mathbb E \left [ \sum_{t=0}^{T} r_t \right ],
        \end{equation}
        where $\tilde{r}_t = R(\tilde{x}_t, \tilde{a}_t)$,  $\tilde{x}_t \sim p(x_t | x_{t-1}, a)$, $\tilde{x}_0 \sim p(x_0)$ and $\tilde{a}_t \sim \pi(a_t | x_t)$. The expectation is computed over all of these sampling processes.

    \subsection{Policy gradients and REINFORCE}
        One approach to obtain the optimal policies is to use the PG~\cite{sutton2000policy} to compute the gradient of the learning objective with respect to the policy. Compared to other approaches such as \textit{value-based learning} \cite{sutton2018reinforcement}, PG can be applied more generally; the only requirement is that the system must output probabilities for the actions it chooses. If the policy $\pi_{\params}$ is parametrized with parameters $\params$ (e.g., parameters of a neural network), the PG can be used to update parameters as follows:\footnote{While \cite{sutton2000policy} derives this for state-action values (not covered here), this also works for the sum of episodic returns (used here).}
        %
        %\begin{equation}
        %    \label{eq:pg}
        %    \nabla_\theta \mathbb{E} \left [ \sum_{t=0}^T r_t \right ] = \mathbb{E} \left [ \sum_{t=0}^T \nabla_\theta \log \pi(a_t | x_t) \left ( \sum_{t=0}^T r_t \right )  \right ] ~ , 
        %\end{equation}
        %
        %\textcolor{blue}{
        %(JY comment: again, relationship between $\params$ and equation 8 is not intuitive. How about adding one equation below and showing them like below for introductory readers? Note that I added $\params$ to $\pi$, that is, $\pi_{\params}$. %\textit{Using the so-called log-derivative trick, the PG updates the parameters as follows:} 
        %\begin{align}
        %    \params' &= \params + \alpha \nabla_\theta \mathbb{E} \left [ \sum_{t=0}^T r_t \right ] \\ 
        %    \label{eq:pg} \nabla_\theta \mathbb{E} \left [ \sum_{t=0}^T r_t \right ] &= \mathbb{E} \left [\nabla_\theta \log \pi_{\params}(a_t | x_t) \left ( \sum_{t=0}^T r_t \right )  \right ] ~ , 
        %\end{align}
            \begin{equation}
            \params' = \params + \alpha \nabla_\theta \mathbb{E} \left [ \sum_{t=0}^T r_t \right ], 
            \end{equation}
            where the gradient of the expectation is 
            \begin{equation}
                \label{eq:pg} \nabla_\theta \mathbb{E} \left [ \sum_{t=0}^T r_t \right ] = \mathbb{E} \left [\nabla_\theta \log \pi_{\params}(a_t | x_t) \left ( \sum_{t=0}^T r_t \right )  \right ] ~ ,
            \end{equation}
        where $\nabla_\theta \bullet$ denotes the vector of all partial derivatives of $\bullet$ with respect to $\theta$, or the gradient. The gradient of the objective function can be increased by moving the policy's parameters according to the gradient's direction in a small step (a \textit{gradient ascent} step).
        
        Intuitively, the right side of PG \eqref{eq:pg} suggests that to increase the expected return, one should encourage the policy to take actions that lead to a high sum of returns and discourage actions that lead to a negative sum of rewards. This becomes even more apparent in our ASV setting, where we only have \textit{one} step and \textit{one} reward---the system sees a single state (the input), determines an action, and gets rewarded. If we set the reward for the correct decision to be $r = 1$, the PG suggests that we should encourage the ASV system to take that same action more often. Likewise, if the decision was wrong and we set the reward to be negative $r = -1$, the PG suggests that we should reduce the probability of that action to improve our reward.
        
        One of the earliest RL algorithms using PG is REINFORCE~\cite{williams1992simple}, which iteratively updates policy by performing two steps:
        \begin{enumerate}
            \item Run a policy in the environment for multiple episodes.
            \item Estimate PG with collected data and update the policy.
        \end{enumerate}
        If the completed episodes have a high/positive sum of rewards, the actions the policy took will be encouraged; likewise, actions related to bad episodes will be discouraged. This estimation of the PG has been shown to have a zero bias but a high variance---PG does provide the correct gradient at expectation, but in practice, any single sampled estimation has a large error. Reducing this variance and improving PG-based methods is an ongoing area of research (e.g., \cite{ppo} and \cite{gae}).

\section{Optimizing tandem systems with reinforcement learning}
\label{sec:method}
    Building upon the background information provided in Section \ref{sec:rl}, we now describe how to use RL to optimize tandem systems composed of ASV and CM systems. We also compare the described method to alternative solutions. 

    \subsection{Tandem systems as Markov decision processes (MDPs)}
        \label{sec:reinforce}
\begin{figure}[t]
        \centering
            \includegraphics[width=0.6\columnwidth]{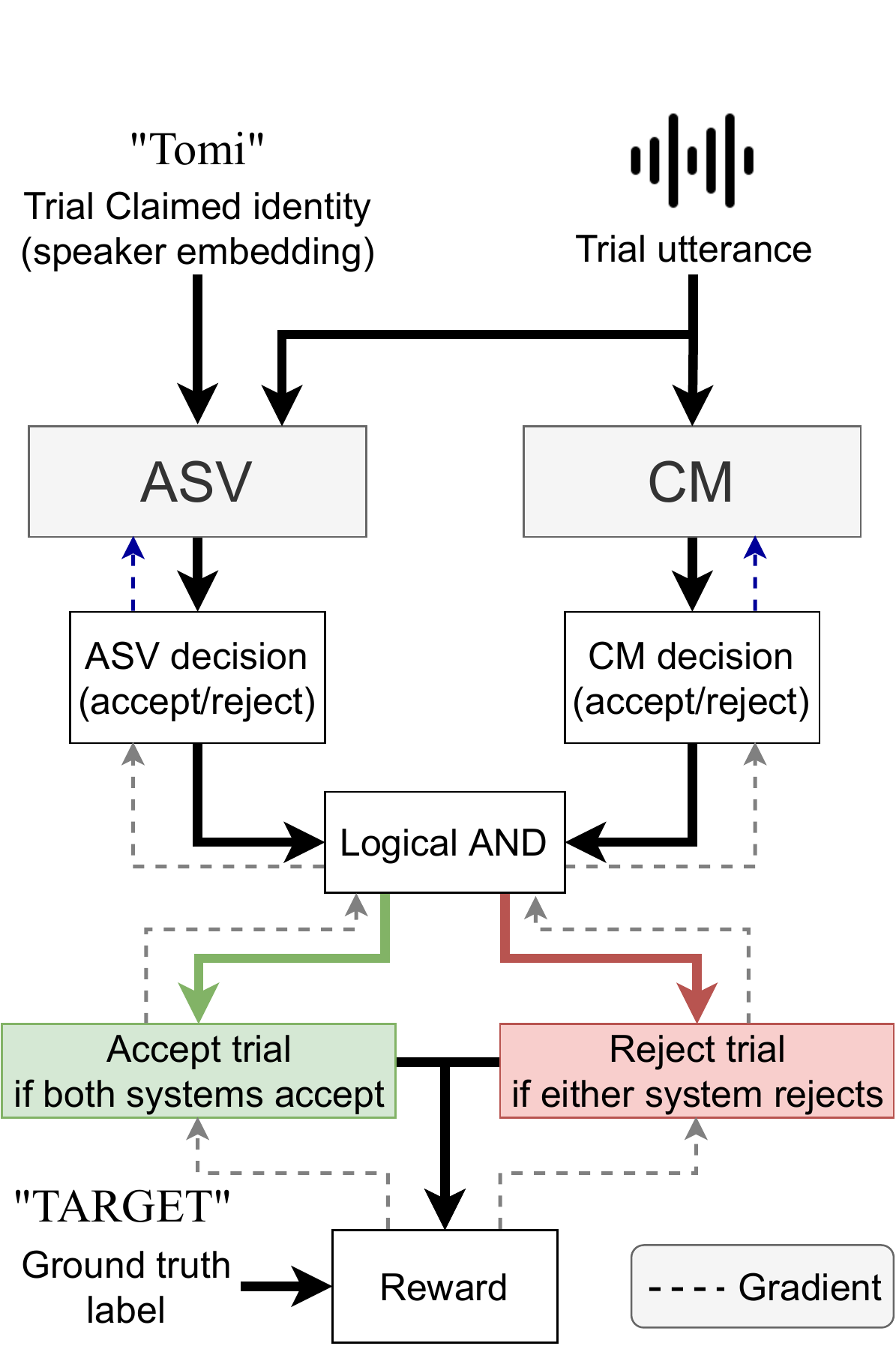}
            \caption{\revised{Illustration of the parallel ASV and CM tandem system. Dashed arrows represent the backpropagation of errors (gradients). Blue dashed lines represent where the PG theorem \eqref{eq:pg} is used.}}
            \label{fig:parallel-system}
        \end{figure}

        The training procedure is outlined in Algorithm \ref{algo:reinforce}, \revised{and the flow of information is illustrated in Figure \ref{fig:parallel-system}}. The tandem system consisting of ASV and CM systems can be considered a simple MDP with only one transition and two actions: \texttt{ACCEPT} or \texttt{REJECT}. In each episode, both systems receive their respective inputs and determine an action via stochastic sampling. \revised{In practice, this process is performed by picking a uniform random number from the interval $[0,1]$ and checking if it is lower than or equal to the sigmoided output value; if so, the decision is \texttt{ACCEPT}; otherwise, \texttt{REJECT}. If the sigmoided output value is close to one, then the decision is almost certainly \texttt{ACCEPT}. The final decision is determined by computing the logical-and of the two system decisions.}
        \begin{algorithm}[t]
        \caption{Optimizing a tandem system with REINFORCE.}
        \label{algo:reinforce}
        \begin{algorithmic}
        
        \STATE \textbf{Input:} Pre-trained ASV $\pi_\text{asv}$ and CM $\pi_\text{cm}$ with combined parameters $\params$, dataset $D$, reward function $R$, mini-batch size $B$
        \WHILE {training}
          \STATE Initialize loss $\mathcal L \leftarrow 0$ 
          \FOR {$i \in \{1, 2, \ldots B\}$}
            \STATE //\revised{Sample one trial with true class labels $y_\text{asv}$ and  $y_\text{cm}$}
            \STATE $x_\text{asv}, x_\text{cm}, y_\text{asv}, y_\text{cm} \sim D$ 
            \STATE //Sample actions (binary decisions).
            \STATE $a_\text{asv} \sim \pi_\text{asv}(x_\text{asv})$ 
            \STATE $a_\text{cm} \sim \pi_\text{cm}(x_\text{cm})$ 
            \STATE //Combine for tandem action. 
            \STATE $a_\text{tandem} \leftarrow a_\text{asv} \land  a_\text{cm}$ 
            \STATE //Compute probability of the tandem action. 
            \STATE \revised{$p_\text{tandem} \leftarrow
            \begin{cases}
                \pi_\text{asv}(x_\text{asv}) \pi_\text{cm}(x_\text{cm}),&a_\text{tandem} = 1 \\
                1 - \pi_\text{asv}(x_\text{asv}) \pi_\text{cm}(x_\text{cm}),&a_\text{tandem} = 0
            \end{cases}$} 
            \STATE //Compute reward. 
            \STATE $r \leftarrow R(a_\text{tandem}, y_\text{asv}, y_\text{cm})$ 
            \STATE //Accumulate policy gradient losses. 
            \STATE $\mathcal L \leftarrow \mathcal L + \frac{1}{B} (\log p_\text{tandem}) r$ 
           \ENDFOR
           \STATE //Backpropagate to obtain gradient $\nabla_{\params} \mathcal L$ and update parameters $\params$ with gradient ascent.
        \ENDWHILE
        \end{algorithmic}
        \end{algorithm}
        
        Finally, the tandem system is rewarded if it correctly accepted or rejected the trial. We then compute the PG loss \eqref{eq:pg} by averaging loss over a mini-batch and update \textit{all} parameters of both systems with a gradient ascent step. We use \texttt{PyTorch} library \cite{pytorch} and its automatic differentiation to compute gradients throughout the ASV and CM systems.

        While this solution is general and can be applied to different systems, two aspects require more in-depth study: how to select an action and what the reward function should be. Picking the action is especially relevant to our scope of ASV and CM systems, as these systems are normally trained to output raw scores rather than probabilities. For the sake of generalization, we assume that the scores are scalars, where a higher value reflects more evidence towards accepting the trial. The reward function, on the other hand, defines the objective function we ultimately wish to maximize (or a loss function we wish to minimize).

    \subsection{Mapping system scores to decision probabilities with calibration}
        \label{sec:calibration}
        \begin{figure}[t]
        \centering
            \includegraphics[width=1\columnwidth]{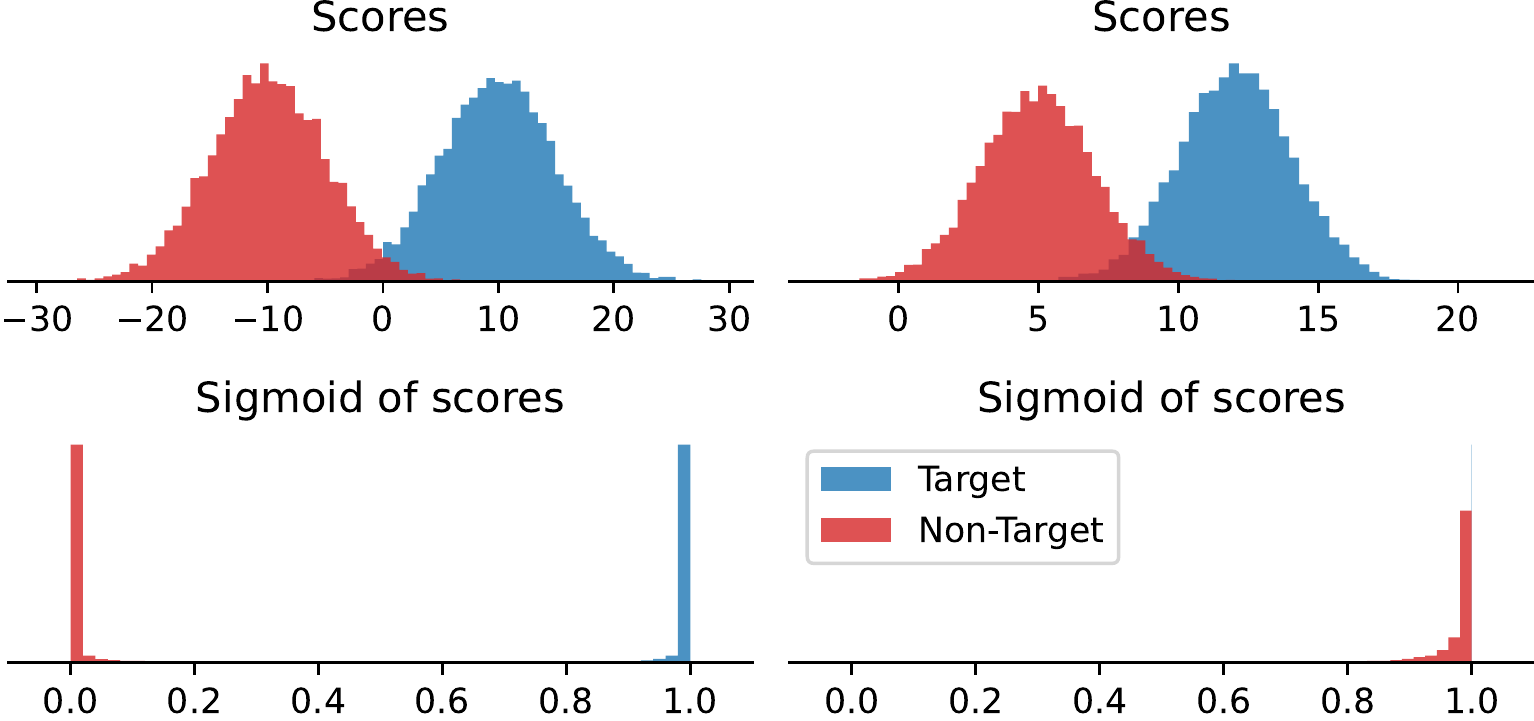}
            \caption{Two illustrations of how sigmoiding scores to produce action probabilities may fail. On the left, the two distributions are assigned to extremes of the probability range, and there will be no exploration. On the right, both distributions are assigned to the same action, and non-target samples would be accepted.}
            \label{fig:score-sigmoiding}
        \end{figure}
        
        %\textcolor{blue}{(JY comment: since this section describes two topics, decision probability and calibration, why don't you describe them as separated paragraphs?)}
        
        By default, the outputs of the system are raw detection scores $s$. To turn these into action probabilities, we could apply the sigmoid function
        \begin{equation}
            \label{eq:sigmoid}
            \sigma(s) = \frac{1}{1 + e^{-s}} ,
        \end{equation}
        where $s$ is a score from either an ASV or CM system and $\sigma$ is the sigmoid function. However, depending on the range of the scores, this design may result in degenerate action probabilities (see Figure~\ref{fig:score-sigmoiding} for a demonstration).
        
        Fortunately, a related problem has been studied in the form of \textit{score calibration} \cite{brummer2006application, brummer2010measuring}. After calibration, the output scores can be treated as \textit{log-likelihood ratios} (LLRs) of two hypotheses~\cite{brummer2006application}
        \begin{equation}
            s = \text{LLR} = \log \frac{p(x | H_s)}{p(x | H_d)} ,
        \end{equation}
        where, for ASV, $H_s$ is the hypothesis that two utterances come from the same speaker and $H_d$ is the hypothesis that utterances come from different speakers; for CM, $H_s$ is the hypothesis that a sample is bonafide and $H_d$ is the hypothesis that the sample is a spoof sample. A convenient property of LLR is that we can obtain a posterior probability of either hypothesis with~\cite{ferrer2021speaker}
        \begin{equation}
            p(H_s | d) = \sigma \left ( s + \log \frac{p(H_s)}{p(H_d)} \right ) ,
            \label{eqn:decision-prob}
        \end{equation}
        where $\log \frac{p(H_s)}{p(H_d)}$ is the \textit{prior log-odds} of the two hypotheses. These two priors are set by the system designer according to their best knowledge on how often target (positive) trials occur~\cite{brummer2014comparison}. 
        Using this relationship, we define the posterior probability $p(H_s | x)$ as the probability of choosing the \texttt{ACCEPT} action: 
        \begin{equation}
            \pi_{\params}(a_t | x_t)  = p(H_s | d),
        \end{equation}
        %
        %(JY comment: Since this is a little bit redundant, we may show this alternatively if you want 
        %\begin{equation}
        %    \pi_{\params}(a_t | x_t)  = p(H_a | d) = \sigma \left ( s + \log \frac{p(H_a)}{p(H_r)} \right ) ,
        %\end{equation}
        %)
        %}
        
        To calibrate the systems, we use parametric discriminative calibration, where we transform scores with an affine transformation $s' = as + b$, where the calibration parameters $(a, b)$ are obtained via training~\cite{brummer2014comparison} and $s'$ is used in \eqref{eqn:decision-prob} instead of $s$. Following \cite{brummer2014comparison} and \cite{garciaromero2020magneto}, we minimize the loss function to obtain $(a, b)$:
        \begin{align}
            \mathcal L_\text{c}(a, b) = \frac{P(H_s)}{|\mathcal P|} \sum_{x \in \mathcal P} \log \left ( 1 + e^{p(H_s | x)} \right ) + \\ \frac{P(H_d)}{|\mathcal N|} \sum_{x \in \mathcal N} \log \left ( 1 + e^{-p(H_d | x)} \right ) ,
        \end{align}
        where $\mathcal P$ and $\mathcal N$ are the sets of positive (same speaker) and negative (different speaker) trials, respectively. 

    \subsection{Optimizing for the t-DCF}
        \label{sec:tdcf-reward}
        Above we described a simple reward function of $r=1$ for correct decisions and $r=-1$ for incorrect decisions. However, while maximizing this reward function would improve the tandem system, it will not optimize the t-DCF function itself. A convenient property of PG \eqref{eq:pg} is that it maximizes any reward function on the right-hand side. Therefore, if we substitute the negative of our t-DCF function as a reward function, we can then directly optimize the t-DCF: maximizing the negative of the t-DCF leads to minimizing the t-DCF.
        
        \revised{Recall that the t-DCF \eqref{eq:new_tdcf} consist of three separate error terms/scenarios: the tandem system rejects a target speaker (miss), the system accepts a non-target speaker (false accept, or fa) or the system accepts a spoof trial (fa, spoof). If we compute t-DCF for a single trial, we will have exactly one of these errors or a true positive/negative. We can describe the cost of an individual trial as follows:}
        
        \begin{equation}
            \label{eq:tdcf-reward}
            \text{t-DCF}_\text{single} = \begin{cases}
                \multirow{2}{*}{$C_{\text{miss}} \cdot \rho_{\text{tar}},$} 
                    & a_{\text{tandem}} = \texttt{REJECT} \\ 
                    & \revised{y_\text{asv} = \texttt{TARGET},} \\
                    & \revised{y_\text{cm} = \texttt{BONAFIDE}}, \\ \\
                \multirow{3}{*}{$C_{\text{fa}} \cdot \rho_\text{non},$} 
                    & a_{\text{tandem}} = \texttt{ACCEPT} \\
                    & \revised{y_{\text{asv}} = \texttt{NONTARGET}} \\
                    & \revised{y_{\text{cm}} = \texttt{BONAFIDE}}, \\ \\
                \multirow{2}{*}{$C_{\text{fa,spoof}} \cdot \rho_\text{spoof},$}
                    & a_{\text{tandem}} = \texttt{ACCEPT} \\ 
                    & \revised{y_\text{asv} = \texttt{TARGET}}, \\
                    & \revised{y_\text{cm} = \texttt{SPOOF}}, \\ \\
                0, & a_{\text{tandem}} = y_{\text{tandem}}
            \end{cases}
        \end{equation}
        where $y_\text{asv}$ and $y_\text{cm}$ are true labels for the ASV and CM systems, respectively, and $y_{\text{tandem}}$ is the 'logical-and' of these true labels. \revised{Actions $a$ are the system outputs. We assume that all spoof trials aim to mimic the target speaker.} We take the negative of the $\text{t-DCF}_\text{single}$ to construct the reward function for optimizing the t-DCF. \revised{Much like the t-DCF metric used for evaluation \eqref{eq:new_tdcf}, a core aspect of this reward function is that different error situations are weighted differently according to their impact (cost) and prevalence (prior terms). Even if the tandem optimization were not able to improve the discriminative performance of the systems, weighing errors according to the t-DCF may bias the systems in the desired direction (e.g., assigning more cost to incorrectly accepting spoofing attacks biases the CM system to reject borderline trials).}

\section{Optimizing tandem systems with differentiable metrics}
\label{sec:method2}
    An alternative approach to RL is to design a differentiable (\say{soft}) version of the t-DCF, which then can be minimized by gradient descent methods. This solution has been used with other metrics \cite{kukanov2016deep, gao2006maximal, sizov2017direct}, including DCF~\cite{mingote2019optimization}, but not with t-DCF. Given its success with other metrics, we extend the idea of soft DCF to t-DCF to assess its applicability in tandem optimization.
    
    The single non-differentiable operation in both DCF and t-DCF metrics are the miss $P_\text{miss}$ and false acceptance $P_\text{fa}$ rates, defined as
    \begin{align}
        P_\text{miss} &= \frac{1}{|\mathcal P|} \sum_{x \in \mathcal P} \mathbbm{1}(\pi(x) \leq \tau) \\
        P_\text{fa} &= \frac{1}{|\mathcal N|} \sum_{x \in \mathcal N} \mathbbm{1}(\pi(x) > \tau) ,
    \end{align}
    where $\pi$ is the detection system, $\tau$ is the decision threshold and $\mathbbm{1}$ is the indicator function that outputs 1 if the condition is true and 0 otherwise. This indicator function is non-differentiable.
    
    The authors in \cite{mingote2019optimization} proposed to soften these rates by applying the following approximations
    \begin{align}
        \label{eq:soft-rates}
        \hat P_\text{miss} &= \frac{1}{|\mathcal P|} \sum_{x \in \mathcal P} \sigma(\tau - \pi(x)) \\
        \hat P_\text{fa} &= \frac{1}{|\mathcal N|} \sum_{x \in \mathcal N} \sigma(\pi(x) - \tau) ,
    \end{align}
    where $\sigma$ is the sigmoid function \eqref{eq:sigmoid}. Essentially, the failure rates are approximated by their distance from the threshold. The soft DCF derived using these rates can then be minimized, and to minimize for any available threshold, the authors proposed optimizing for the threshold $\tau$ as well as the parameters of the detection system.
    
    Following this approach, we derive a soft t-DCF by applying \eqref{eq:soft-rates} to the error rates of \eqref{eq:new_tdcf}. Optimization can therefore be applied to tune the systems and scores for the priors and cost weights set by the evaluation metric.

    % Anssi: Feels like this should be earlier in the paper, if anywhere?
    % \subsection{Related work}
        % REINFORCE, or also called \say{score function estimator} \cite{jang2017categorical}, is no stranger to supervised learning. It has been used to \textit{e.g.} compute gradient through loss metric that requires generating a sequence of characters \cite{yu2017seqgan}. The version of REINFORCE used in this work is the simplest version, but multiple alternatives have been proposed over the years with stabler learning, summarized by \cite{jang2017categorical} along with their proposed method (differentiable softmax operation). 
        
        % The reinforcement learning model used here for the tandem ASV-CM setup and training is not the only one available, either. One could, for example, model the situation as a \textit{multi-agent} setup with multiple individual policies \cite{boutilier1996planning}, or use the \textit{cascaded} setup where CM first has to accept the trial before passing it to the ASV \cite{kinnunen2018t}. Our focus is on studying the applicability of REINFORCE on optimizing such tandem systems in terms of t-DCF, and thus opt for the simplest approach.

\section{Experimental setup}

    Here we cover the datasets, features (front-end), neural network architectures, and training procedures used in the experiments to evaluate the above solutions. Before tandem training, we pre-train the ASV and CM systems separately. We repeat all the experiments three times and report the average results to avoid drawing conclusions influenced by noise~\cite{henderson2018deep, wang2021comparative}. Such noise may stem from, for example, random initialization of the network, the sampling of data, and the sampling of actions in the REINFORCE method.

    \subsection{Datasets}
        
        \begin{table}[t]
            \centering
            \caption{Statistics of the ASVspoof19 dataset (logical access) for training spoofing countermeasure systems, with the number of speakers by gender.}
            \label{table:asvspoof_stats}
            \begin{tabular}{lcccc}
            \toprule
            \textbf{Partition}& \textbf{Male} & \textbf{Female} & \textbf{Bonafide} & \textbf{Spoof} \\ \midrule
            Train    & 8             & 12      & 2,580      & 22,800 \\
            Development      & 8             & 12      & 2,548      & 22,296 \\
            Evaluation     & 30            & 37      & 7,355      & 63,882 \\
            \bottomrule
            \end{tabular}
        \end{table}
        
        The main dataset used in this work is the ASVspoof19 dataset (logical access, or LA, scenario) \cite{todisco2019asvspoof}, which provides labels for speaker verification and spoof samples generated using different techniques. We use the ASVspoof19 dataset to train the initial countermeasures and to perform the tandem training. Statistics of the portion used for training independent ASV and CM are summarized in Table \ref{table:asvspoof_stats}.

        As ASVspoof19 does not contain enough data on different speakers for training a reliable speaker verification system, we use the VoxCeleb2 dataset~\cite{chung2018voxceleb2} to pre-train it. Specifically, we use the development part of the authors' protocol, which contains 5,994 unique speakers and 1,092,009 utterances. \revised{The whole VoxCeleb2 dataset was extracted from 150,480 unique videos (2,442 hours) with an average individual utterance length of 7.8 seconds; face recognition was used to capture speech of the persons of interest (celebrities). In this dataset, 61\% of the speakers are male and 39\%, female. Roughly 39\% of the speakers are from the USA or UK by nationality~\cite{chung2018voxceleb2}. We do not use evaluation sets of the VoxCeleb2 in this work.}
    
    \subsection{Evaluation metrics}
        \label{sec:evaluation-metrics}
        We evaluate the tandem systems using the normalized minimum t-DCF with the same parameters and protocol as in the ASVspoof19 challenge \cite{todisco2019asvspoof}. This approach uses the ASV-constrained form of the t-DCF, where we fix the threshold of the ASV system to its \textit{equal error rate} (EER) point on target-non-target discrimination and then sweep over different thresholds for the CM system to obtain a minimal, normalized t-DCF. Individual ASV and CM systems are evaluated on their respective tasks with EER (ASV on target-non-target discrimination; CM on bonafide-spoof discrimination). Unless otherwise noted, all the t-DCF and EER values are computed using the ASV protocols provided by the ASVspoof19 dataset (trials in \texttt{ASVspoof2019.LA.asv.dev.gi.trl.txt}\footnote{Available in the ASVspoof19 dataset at \url{https://datashare.is.ed.ac.uk/handle/10283/3336}}), including the performance of the CM system.

    % \begin{figure}[t]
    %     \includegraphics[width=\columnwidth]{material/models.pdf}
    %     \figsqueezevspace
    %     \caption{Neural network architectures for ASV (left) and CM (right) models. All hidden layers use ReLU activations. Blue boxes represent ``siamese" modules and green ``discriminator" modules. ASV system is a classification back-end with fixed embeddings, while CM system contains both front and back-ends. Time dilated layers follow the notation of \cite{peddinti2015time}, and CM model is a lighter version of X-vector model described in \cite{snyder2018x}.}
    %     \label{fig:models}
    % \end{figure}

    \subsection{Speaker verification system (ASV)}
        \label{sec:asv}
        The ASV system is inspired by the results of x-vector (or r-vector) systems \cite{garciaromero2020magneto} (MagNetO) and \cite{zeinali2019but} (BUT-system), both of which have achieved state-of-the-art results in multiple ASV corpora. Due to memory limitations, we use the network architecture of the BUT-system~\cite{zeinali2019but} with the short-segment training setup from MagNetO~\cite{garciaromero2020magneto}, with the following modifications. First, we set the batch size to 32 (instead of 256) to accommodate our memory constraints. To compensate, we train for 200,000 steps (instead of 150,000). We use standard cross-entropy loss with the Adam optimizer to train the system with an initial learning rate of 0.001 and halve the learning rate every 10,000 training steps up to 100,000 training steps. We also use a small L2 regularization weight of $10^{-5}$. These settings were selected based on preliminary results with cosine scoring on the evaluation set of VoxCeleb2.
        
        During training, each utterance is augmented with one of the following: music, babble, reverberation, or noise. We use impulse responses from the AIR RIR database \cite{jeub2009binaural} for reverberation and the MUSAN database \cite{snyder2015musan} for speech, music, and noise sources. After augmentation, we cut out a random four-second segment from the utterance or discard the utterance completely if the utterance is shorter than four seconds. Log-mel filterbank energies are used as features (80 features per frame, 25ms window length, 10ms window hop), post-processed with sliding-window mean-normalization.

        For the ASV back-end we chose to use \textit{neural PLDA} (NPLDA)~\cite{ramoji2020nplda}, which we train using the same VoxCeleb2 dataset with 100 epochs. \revised{The non-neural alternative, PLDA~\cite{prince2007probabilistic, sizov2014unifying}, has been a successful back-end in ASVSpoof19 dataset~\cite{todisco2019asvspoof}. Compared to PLDA, NPLDA is suitable for gradient updates which we need for our tandem training.} Each epoch consists of 1,000 mini-batches, each with 2,048 random pairs of utterances (and a balanced number of target and non-target pairs). The NPLDA model is trained with cross-entropy and L2 regularization (weight $10^{-5}$) losses using the Adam optimizer with a learning rate of 0.0001, which is halved every 20 epochs. \revised{All parameters were taken from the source code of the original NPLDA publication~\cite{ramoji2020nplda}, apart from training length. For training length. We increased training length until training loss converged.}
        
        After training the system using the VoxCeleb2 dataset, we adapt the NPLDA to ASVspoof19 corpus ASV part (bonafide trials from \texttt{ASVspoof2019.LA.cm.trn.txt}) by finetuning its parameters for one epoch, using the same settings, which we found to improve performance without overfitting with this small dataset. This approach reduced the EER from an average of 10.1\% to 5.8\% in the ASVspoof19 evaluation set. \revised{Training longer than one epoch worsened the results.}

    \subsection{Spoofing countermeasure (CM) system}
        For the CM system, we use the source code and a system similar to the one described in \cite{wang2021comparative}, where a \textit{light convolutional neural network} (LCNN) network of \cite{lavrentyeva2019stc} is used with \textit{linear frequency cepstral coefficient} (LFCC) features. \revised{We chose this system based on its high accuracy with the ASVSpoof19 dataset at the time of these experiments~\cite{wang2021comparative} and for its neural-based architecture, which we require for the gradient updates in the tandem training}. The LFCC features follow the baseline recipe presented in the ASVspoof19 \cite{todisco2019asvspoof}, with 512-length fast Fourier transform computed with 20ms windows and 10ms hops, processed with 20 linearly spaced filterbanks, along with delta and delta-delta features. Every utterance is cropped or padded to a fixed size of 750 feature vectors. Shorter utterances are padded with zeros from the end, and longer utterances are randomly cropped to fit 750 feature vectors. The model outputs a single sigmoid scalar and is trained on the dataset using cross-entropy. For the description of the \texttt{LCNN-trim-pad (sigmoid)} system in \cite{wang2021comparative}.

    \subsection{Tandem training and baseline}
        % \begin{figure}[t]
        % \centering
        %     \includegraphics[width=1\columnwidth]{material/methods.pdf}
        %     \caption{\todo{Update with different methods we compare} Methods for optimizing tandem systems explored in this work, along with the baseline where the systems are only trained separately. ``ASV+CM dataset" refers to dataset with only tandem accept/reject labels. Dashed lines represent flow of training information (i.e. backpropagation).}
        %     \label{fig:methods}
        % \end{figure}
        
        After developing the pre-trained systems (See \say{Initial} results in Table \ref{table:tdcf_results}), we aim to optimize them using the techniques described in Sections~\ref{sec:method} and \ref{sec:method2}. In all cases, we optimize the parameters of the NPLDA back-end of the ASV system and all parameters of the CM system. We only train the back-end of the ASV system due to memory limitations. As a baseline solution, we use the same training regimen to finetune the pre-trained systems separately using the same data. We emphasize that no tandem training is used in this setup---we merely use the same data to further update the models for a fair comparison against the proposed methods.

        For tandem optimization, we use the trial list from ASVspoof19's development list, (\texttt{ASVspoof2019.LA.asv.dev.gi.trl.txt}). This file list contains 10 different speakers with 2,548 bonafide and 22,296 spoof samples. To avoid catastrophic forgetting and overfitting to the small number of samples of this dataset, we use standard stochastic gradient descent (without momentum) with a learning rate of 0.0001 and mini-batch size of 64, with a total of five epochs of training over the entire dataset. Each item in a mini-batch has an equal sampling probability of being either a target, a non-target, or a spoof trial.
        
        In total, we compare six different methods. We use the same t-DCF cost and prior parameters for these training methods as with our evaluation metric. These six methods are as follows:
        \begin{enumerate}
            \item Finetune.
            \item REINFORCE without calibration (Section \ref{sec:reinforce}).
            \item REINFORCE with calibration (Section \ref{sec:calibration}).
            \item REINFORCE with t-DCF reward and without calibration (Section \ref{sec:tdcf-reward}).
            \item REINFORCE with t-DCF reward and calibration (Sections \ref{sec:calibration} and \ref{sec:tdcf-reward}).
            \item Soft t-DCF (Section \ref{sec:method2}).
        \end{enumerate}
        
        % \textbf{REINFORCE.}
        %     \todo{Update this?}
        %     For the REINFORCE method described in Section \ref{sec:method}, we use same parameters for reward as used for the t-DCF evaluation: $C_\text{miss} = 1$, $C_\text{fa} = C_\text{fa,spoof} = 10$, $\rho_\text{tar} = 0.95 \cdot 0.99, \rho_\text{non} = 0.95 \cdot 0.01$ and $\rho_\text{spoof} = 0.05$.

\begin{table}[t]
        \centering
        \footnotesize
        \caption{\revised{Results of tandem training averaged over three repetitions, evaluated on the ASVspoof19 dataset. The development dataset is used for tandem training. ``Initial" results are results before tandem training. Best results are in bold per consecutive column, with grey values representing worse performance after tandem training than initially. See Section~\ref{sec:excluded-attacks} for further details on outlier spoofing attacks. Outlier spoofing attacks do not affect development set results.}}
        \label{table:tdcf_results}
        \begin{tabular}{lcccccc}
        \toprule
\textbf{}          & \multicolumn{3}{c}{\textbf{Development}}            & \multicolumn{3}{c}{\textbf{Evaluation}}             \\
\textbf{Method}    & \textbf{\begin{tabular}[c]{@{}c@{}}ASV\\EER\end{tabular}} & \textbf{\begin{tabular}[c]{@{}c@{}}CM\\EER\end{tabular}} & \textbf{t-DCF} & \textbf{\begin{tabular}[c]{@{}c@{}}ASV\\EER\end{tabular}} & \textbf{\begin{tabular}[c]{@{}c@{}}CM\\EER\end{tabular}} & \textbf{t-DCF} \\ 
Initial            & 7.86               & 0.16                       & .0089         & 5.77             & 3.22          & .0768 \\ \midrule
Finetune           & 6.07               & \textbf{0.12}              & \textbf{.0022}& \worse{6.50}      & \worse{3.59}   & \textbf{.0765} \\
REINFORCE          & \textbf{5.92}      & \worse{0.22}                & \worse{.0117}  & \worse{5.78}      & \textbf{3.30} & \worse{.0800} \\
REI. Calib.        & 6.78               & \worse{0.25}                & \worse{.0122}  & \textbf{5.76}    & \worse{3.40}   & \worse{.0835} \\
REI. t-DCF         & 7.19               & \worse{0.20}                & .0055         & \worse{5.80}      & \worse{3.51}   & \worse{.0781} \\
REI. Calib. t-DCF  & 7.67               & \worse{0.19}                & .0055         & 5.77             & \worse{3.52}   & \worse{.0782} \\
Soft t-DCF         & 7.22               & \worse{0.17}                & .0041         & \worse{5.79}      & \worse{3.66}   & \worse{.0782} \\ \midrule
\multicolumn{7}{l}{\textbf{Results without outlier spoofing attacks (A17--A19)}} \\
Initial             &-                &-                &-               &5.77                &0.22                &.0066               \\ \midrule
Finetune            &-                &-                &-               &\worse{6.5}         &\worse{0.23}        &\worse{.0076}       \\
REINFORCE           &-                &-        &-       &\worse{5.78}        &0.19                &.0057               \\
REI. Calib.    &-                &-        &-       &\textbf{5.76}                &\textbf{0.18}                &.0056               \\
REI. t-DCF     &-                &-         &-               &\worse{5.8}         &\textbf{0.18}                &\textbf{.0054}               \\
REI. Calib. t-DCF&-                &-        &-               &5.77                &0.19                &\textbf{.0054}               \\
Soft t-DCF          &-                &-        &-               &\worse{5.79}        &0.19                &.0057               \\
\bottomrule
\end{tabular}
    \end{table}

\begin{figure*}[t]
    \centering
    \includegraphics[width=.98\textwidth]{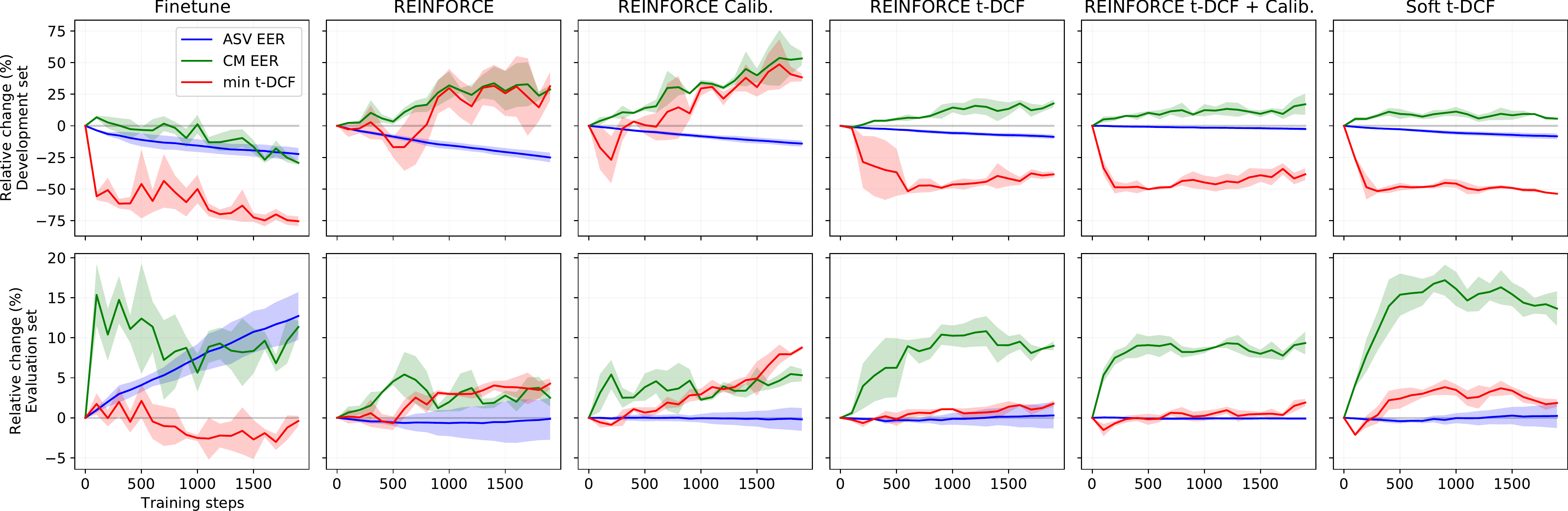}
    \caption{Performance metrics over tandem optimization, evaluated on development (training) and evaluation sets (first and second row, respectively). Y-axis shows the change in the metric relative to the model before tandem optimization. Curves are averaged over three experiments, with shaded regions representing standard deviation.}
    \label{fig:learning_curves}
\end{figure*}

\begin{figure*}[t]
    \centering
    \includegraphics[width=.98\textwidth]{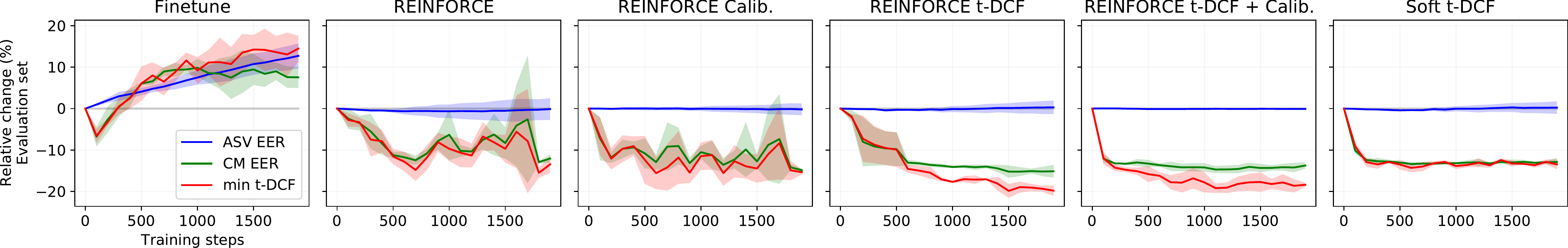}
    \caption{Tandem training learning curves in the evaluation set with three spoofing attack systems removed (A17, A18, A19). Solid lines are averaged over three repetitions, with shaded regions representing the standard deviation.}
    \label{fig:learning_curves_excluded}
\end{figure*}

\section{Results}
    We report results in two ways: tables of raw metrics are provided in Table~\ref{table:tdcf_results}, and learning curves of relative changes are summarized in Figure~\ref{fig:learning_curves}. The learning curves show the change in a metric as training progresses. This allows us to study how the change in one metric affects others and, importantly, how ASV and CM performance affects t-DCF. 
    
    All optimization groups are able to improve t-DCF in the training set, but not reliably in the evaluation set, \revised{unless we remove outlier spoofing attacks (these results are further discussed in Section~\ref{sec:excluded-attacks})}. Interestingly, while finetuning improves the EER of both ASV and CM in the training set, REINFORCE with t-DCF reward and score calibration only marginally improves ASV EER and worsens CM EER while still improving t-DCF. This finding indicates that tandem training is able to not only optimize the individual systems for better discrimination but also optimize for the operating point induced by t-DCF and its cost function.
    
    In the evaluation set, finetuning reduces the t-DCF while the EER of both ASV and CM systems \textit{increase}. This observation is interesting, as finetuning is not done in tandem, so we would expect any improvement in t-DCF to come from the two systems collaborating. Nonetheless, results indicate that the systems sacrifice per-task performance for better tandem results.
    
    While REINFORCE approaches do not yield positive results, they have one advantage over finetuning: the performance of the ASV system does not get considerably worse. The disparency between the training and evaluation set in the ASV EER suggests that finetuning overfits to the training set, but this is not the case for REINFORCE. Optimizing for soft t-DCF yields similar results to REINFORCE but with more pronounced changes in metrics.

\begin{table*}[t]
        \centering
        \footnotesize
        \setlength{\tabcolsep}{.5em}
        \caption{The CM EERs (\%) after tandem training separated by the spoofing attacks. Values in grey indicate results worse than the initial EER. Systems A01-A06 are included in the development (training) set, while A07-A19 are only in the evaluation set.}
        \label{table:cm-eers}
        \begin{tabular}{lccccccccccccccccccc}
        \toprule
\textbf{Method}     &\textbf{A01}        &\textbf{A02}        &\textbf{A03}        &\textbf{A04}        &\textbf{A05}        &\textbf{A06} \hspace{0.3cm}        &\textbf{A07}    &\textbf{A08}        &\textbf{A09}        &\textbf{A10}        &\textbf{A11}        &\textbf{A12}        &\textbf{A13}        &\textbf{A14}        &\textbf{A15}        &\textbf{A16}        &\textbf{A17}        &\textbf{A18}        &\textbf{A19}        \\
Initial             &0.1                 &0.1                 &0.0                 &0.3                 &0.2                 &0.1 \hspace{0.3cm}  &0.1                 &0.2                 &0.0                 &0.3                 &0.0                 &0.3                 &0.2                 &0.0                 &0.1                 &0.3                 &16.7                &2.4                 &1.1                 \\ \midrule
Finetune            &0.0                 &0.0                 &0.0                 &0.2                 &0.2                 &0.1 \hspace{0.3cm}  &\worse{0.2}          &\worse{0.3}         &0.0                 &\worse{0.3}         &\worse{0.2}         &0.3                 &0.2                 &\worse{0.1}         &\worse{0.2}         &\worse{0.3}         &\worse{19.2}        &1.9                 &\worse{1.2}         \\
REINFORCE           &\worse{0.1}         &0.0                 &0.0                 &\worse{0.3}         &\worse{0.3}         &\worse{0.2} \hspace{0.3cm}           &0.1                 &\worse{0.2}         &0.0                 &\worse{0.3}         &0.0                 &0.2                 &0.2                 &0.0                 &0.1                 &0.3                 &\worse{16.7}        &\worse{2.6}         &\worse{1.3}         \\
REI. Calib.    &\worse{0.1}         &\worse{0.1}         &0.0                 &\worse{0.4}         &\worse{0.3}         &\worse{0.2} \hspace{0.3cm}         &0.1                 &0.2                 &0.0                 &\worse{0.3}         &0.0                 &0.2                 &0.2                 &0.0                 &0.1                 &\worse{0.3}         &\worse{16.9}        &\worse{2.8}         &\worse{1.4}         \\
REI. t-DCF     &0.0                 &0.0                 &0.0                 &0.2                 &\worse{0.3}         &\worse{0.2} \hspace{0.3cm}         &0.1                 &\worse{0.2}         &0.0                 &0.3                 &\worse{0.1}         &0.2                 &0.2                 &\worse{0.1}         &\worse{0.2}         &0.2                 &\worse{18.2}        &2.1                 &\worse{1.4}         \\
REI. Calib. t-DCF&0.0                 &0.0                 &0.0                 &0.2                 &\worse{0.3}         &\worse{0.2} \hspace{0.3cm}         &0.1                  &\worse{0.3}         &0.0                 &0.3                 &\worse{0.1}         &0.2                 &0.2                 &\worse{0.1}         &\worse{0.2}         &0.3                 &\worse{18.3}        &2.1                 &\worse{1.4}         \\
Soft t-DCF          &0.0                 &0.0                 &0.0                 &0.2                 &\worse{0.4}         &\worse{0.2} \hspace{0.3cm}         &0.1                 &\worse{0.3}         &0.0                 &0.3                 &\worse{0.1}         &0.2                 &0.2                 &\worse{0.1}         &\worse{0.2}         &0.3                 &\worse{19.6}        &1.8                 &\worse{1.3}         \\
\bottomrule
\end{tabular}
\end{table*}

\section{Discussion and further analysis}

    The aggregated results \revised{are less than promising}. While we see positive learning in the training set in all cases, this success does not translate to the unseen attacks in the ASVspoof19 dataset. To understand these results better, we perform further analysis by studying results for each spoofing attack, cross-task performance, and optimization of the ASV front-end.
    
    \subsection{Splitting results by spoofing attacks}
        \label{sec:excluded-attacks}
        The ASVspoof19 dataset consists of 19 different attacks for creating the spoof samples---six in the development set and 13 in the evaluation set. The 13 are all \textit{unseen} attacks. Given the variation of CM performance between these attacks, it is instructive to study per-attack performance~\cite{todisco2019asvspoof}. Table~\ref{table:cm-eers} separates the CM EER results by attack. As with the baseline results in ASVspoof19~\cite{todisco2019asvspoof}, most attacks are well detected with low EERs, apart from the last three (A17--A19). A17 is exceptionally difficult to detect. 
        
        \revised{A common feature of these three attacks is that they modify genuine human speech to perform the spoofing attack, while other systems synthesize speech in one way or another. This feature makes it harder for CM systems to identify A17--A19 as spoofing attacks, but these attacks also do not fool the ASV system (i.e., they are less performant as spoofing attacks)~\cite{todisco2019asvspoof}. Specifically, A17 uses waveform filtering on genuine human speech to perform a spoofing attack but only applies small modifications to the input speech. As such, the utterance resembles the source speaker more than the target speaker (low ASV EER) while remaining difficult to detect (high CM EER). By studying the ASV EERs of different spoofing attacks (Appendix A), we find that attacks A17--A19 are weaker than most of the other attacks. A17 has an ASV EER of 7.7\%, which is close to the bonafide ASV EER of 5.77\% (Table \ref{table:tdcf_results}), indicating that most of the error stems from imperfections in the ASV system. The same was observed in the original ASVSpoof19 publication~\cite{todisco2019asvspoof}.} The difficulty in detecting A17--A19 could also be attributed to the length of silences in these attacks~\cite{muller2021speech}, which makes them outliers among all the attacks. In our tandem training, the CM EERs of these three systems increase in these three cases, while in the majority of other systems the EERs decreases.
        
        Figure~\ref{fig:learning_curves_excluded} and the second half of Table~\ref{table:tdcf_results} present the learning curves and results in the evaluation set \textit{without} A17--A19. All methods except finetuning improve results in all metrics. REINFORCE combined with the t-DCF reward yields the greatest improvement in the t-DCF. Interestingly, while soft t-DCF optimization also reduces t-DCF, it quickly stagnates to a value and no further improvement happens. Meanwhile, REINFORCE with the t-DCF reward shows a declining learning curve, indicating that with further training the results could improve. Curiously, while the original ASV and CM systems were not guaranteed to output scores that are calibrated to be LLRs, as discussed in Section~\ref{sec:calibration}, calibrating the scores and using class posterior probabilities did not seem to have a drastic effect on the training results (Figure~\ref{fig:learning_curves_excluded}, third and fifth panel). The use of reward function based on t-DCF costs, however, stabilized the learning and yielded better performance.

\begin{table}[t]
        \centering
        \footnotesize
        \caption{\revised{Results of evaluating ASV system in CM's task, before and after tandem training. All numbers are equal-error rates in percentages.}}
        \label{table:vice-versa-results}
        \begin{tabular}{lcc}
        \toprule
               & Development & Evaluation \\
\textbf{Method}    & \textbf{\begin{tabular}[c]{@{}c@{}}ASV\\in CM\end{tabular}} & \textbf{\begin{tabular}[c]{@{}c@{}}ASV\\in CM\end{tabular}} \\
Initial             &36.57                                          &44.36                             \\ \midrule
Finetune            &\worse{37.70}                                    &\textbf{43.38}                            \\
REINFORCE           &\textbf{35.95}                              &\worse{44.84}                               \\
REI. Calib.    &36.15                              &\worse{44.88}                             \\
REI. t-DCF     &\worse{37.64}                                  &\worse{44.51}                             \\
REI. Calib. t-DCF&\worse{37.91}                           &\worse{44.47}                             \\
Soft t-DCF          &\worse{38.23}                                &44.24                                   \\
\bottomrule
\end{tabular}
    \end{table}

    \subsection{Analysis of ASV and CM systems in cross-system tasks}
        The goal of tandem optimization is to train the two systems to function better together, with one intuitive benefit being that both systems might learn each other's task---for example, the ASV system could learn to reject spoof samples which the CM system struggles to detect. To this end, Table~\ref{table:vice-versa-results} presents the performance of the ASV system in the CM task. This cross-task performance evaluation is done by taking the scores of the ASV system for a trial but using true spoofing labels to evaluate its CM EER. \revised{We omit results for the CM system's performance in the ASV task, which were 50\% for all cases. This is because the CM system does not receive the claimed identity enrollment utterance as an input, and as such, it can not tell if the input utterance is the target speaker or not.}
        
        The results indicate that the ASV system does not substantially shift towards performing the CM task, especially in the evaluation set. We only see a 1\% EER decrease when using finetuning, which could explain the increased ASV EER of the ASV system in the main results (Figure~\ref{fig:learning_curves}, lower left corner). However, this result cannot be explained by the ASV system learning to perform the CM task, as finetuning does not involve real tandem training and ASV was only trained with ASV labels. Removing the spoofing attack systems A17--A19 and recomputing these numbers did not change the results.

\begin{figure}[t]
    \centering
    \includegraphics[width=.98\columnwidth]{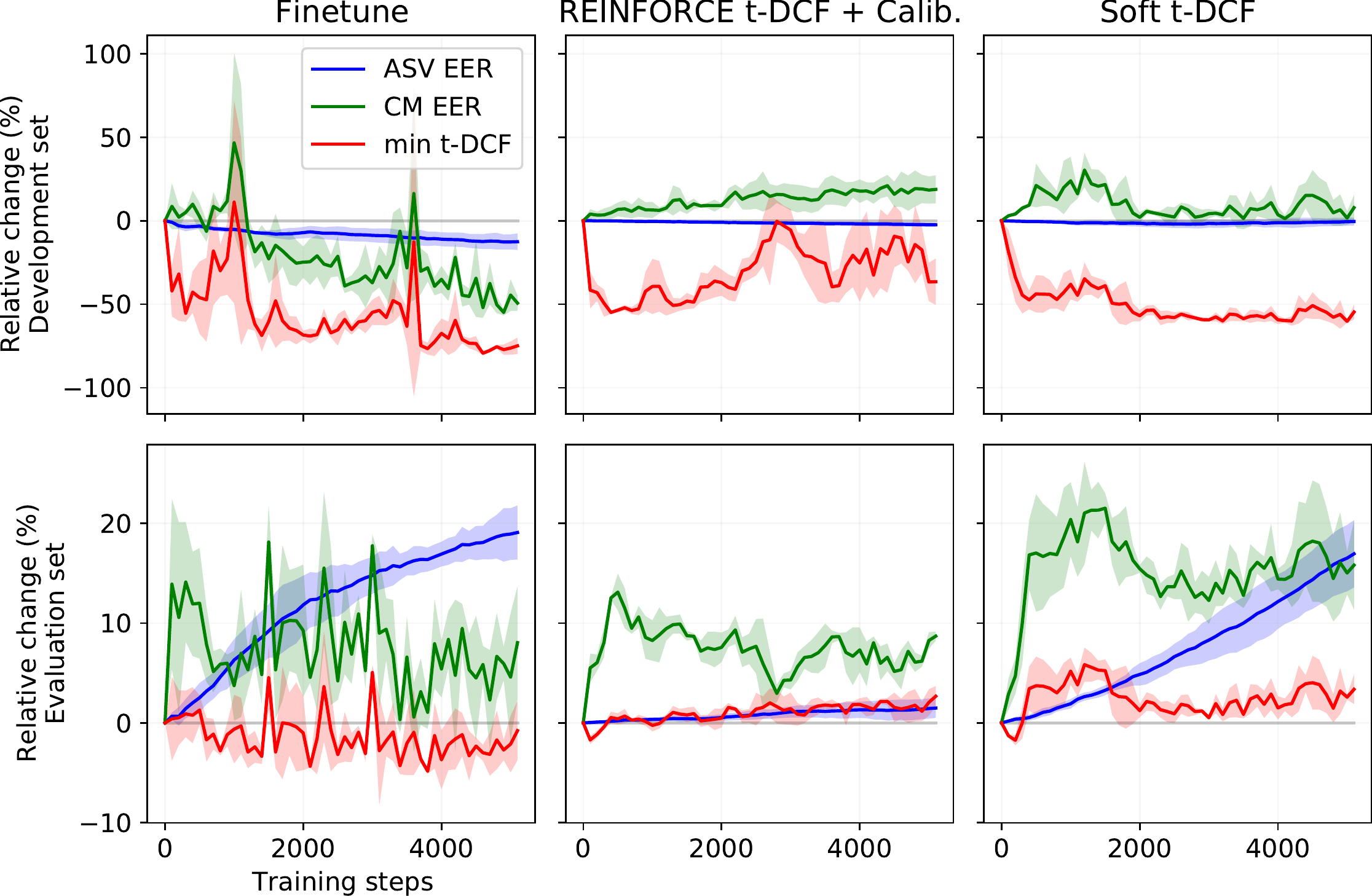}
    \caption{Tandem training learning with ASV utterance embedding (front-end) optimization.}
    \label{fig:learning_curves_frontend}
\end{figure}

\begin{figure}[t]
    \centering
    \includegraphics[width=.98\columnwidth]{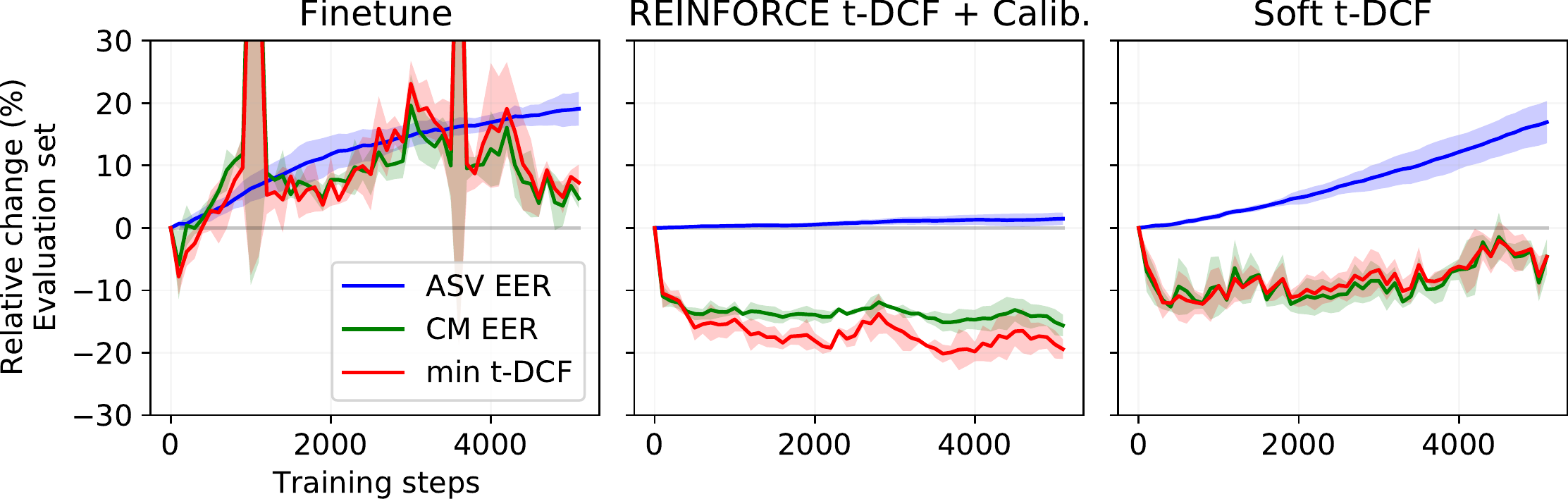}
    \caption{Tandem training learning curves for which we also optimize ASV utterance embeddings, with three spoofing attack systems removed (A17, A18, A19).}
    \label{fig:learning_curves_frontend_excluded}
\end{figure}
    
    \subsection{Training ASV front-end}
        We limited ourselves to optimizing the ASV system's back-end only, as the utterance embedding (front-end, Section \ref{sec:asv}) model was too large to sufficiently fit into memory. The results above indicate that ASV performance does not change considerably, which could be a result of this limitation. To study this finding, we repeat the training procedure but this time we also optimize the ASV utterance embedding model. Due to memory restrictions, we use a batch size of 24 instead of 64, with other settings remaining as above. The policies are thus updated for more steps than in the previous experiments.
        
        The results for all spoofing attacks are included in Figure~\ref{fig:learning_curves_frontend} and for excluded attacks A17--A19, in Figure~\ref{fig:learning_curves_frontend_excluded}. Again, ASV EER either does not change or gets considerably worse. Unlike in the original results, soft t-DCF optimization leads to increased ASV EER in the evaluation set, which seems to affect the tandem system as a whole and result in an increased t-DCF value. REINFORCE yields similar results to earlier findings with no significant change. Note that the CM EER and t-DCF keep improving in REINFORCE training when outlier spoofing attacks are removed, despite the smaller batch size and longer training (Figure~\ref{fig:learning_curves_frontend_excluded}). We hypothesize that the ASV EER does not improve due to lack of training material, in which case it is easy to overfit the training data. Curiously, REINFORCE optimization does not seem to be subject to this flaw.
        
\section{Conclusion}
    ASV and spoofing CM systems are trained separately from each other, yet they are evaluated as one \say{tandem} system. In this work, we evaluated three different approaches for optimizing such tandem systems for spoof-robust speaker verification---finetuning, RL, and a differentiable (soft) tandem evaluation metric. Our results indicate that while traditional finetuning can yield benefits, it tends to overfit the training data. RL-based optimization methods and differentiable evaluation metrics yield solid improvements once outlier systems have been removed (\revised{up to $20\%$ relative improvement}), while finetuning still exhibits overfitting behaviour. These results indicate that the two suggested methods, RL and soft metric, yield the same or better results than naive finetuning of the systems separately and that they should be used for such optimization. RL specifically yielded stable results, regardless of the changes in the training setup, and at least, did not negatively impact the results much like soft metric did occasionally.
    
    Considering that two works in this area have reported similar results (this and the previous work~\cite{kanervisto2020initial}), the tandem optimization of ASV systems combined with a CM system seems promising. Nevertheless, this work should be continued by, for example, performing large-scale experiments with different ASV and CM models. Compared to previous work~\cite{kanervisto2020initial}, this work applied recent state-of-the-art systems and evaluated multiple variations of the tandem optimization methods. Nevertheless, new systems and datasets are continuously presented, and the same tandem optimization experiments could be repeated with these. \revised{Specifically, given the variations in recording conditions and speakers of different datasets, future work should explore the cross-domain performance of tandem training. This could be done by training on one ASVSpoof dataset and evaluating on another, such as ASVSpoof19 and ASVSpoof2021~\cite{yamagishi2021asvspoof}, once the ground truth labels have been publicly shared.} Finally, given the general applicability of the RL approach, we believe that it can be used as a general tool for directly optimizing the evaluation metrics in different domains, but its performance should be evaluated with proper experimentation.

\ifCLASSOPTIONcaptionsoff
  \newpage
\fi



%\bibliographystyle{IEEEtran}
\bibliographystyle{IEEEtran}
\bibliography{refs.bib}

\newpage
\appendices
\section{ASV error rates with spoofing attacks}
Table \ref{table:asv-eers} shows the EERs of the ASV system in discriminating bonafide target trials from spoofing attack trials: if ASV EER is high, the spoofing attack was successful at bypassing the ASV system.
\setcounter{table}{4}
\begin{table*}[]
        \centering
        \footnotesize
        \setlength{\tabcolsep}{.5em}
        \caption{The ASV EERs (\%) before and after tandem training. Non-target trials are spoofing attacks. A high EER indicates that the spoofing attack successfully bypassed the ASV system. Values in grey indicate cases where EER increased during tandem training. The bolded value is the lowest EER over methods (excluding ``Initial"). Systems A01-A06 are included in the development (training) set, while A07-A19 are only in the evaluation set.}
        \label{table:asv-eers}
        \begin{tabular}{lccccccccccccccccccc}
        \toprule
\textbf{Method}     &\textbf{A01}        &\textbf{A02}        &\textbf{A03}        &\textbf{A04}        &\textbf{A05}        &\textbf{A06} \hspace{0.15cm}        &\textbf{A07}        &\textbf{A08}        &\textbf{A09}        &\textbf{A10}        &\textbf{A11}        &\textbf{A12}        &\textbf{A13}        &\textbf{A14}        &\textbf{A15}        &\textbf{A16}        &\textbf{A17}        &\textbf{A18}        &\textbf{A19}        \\
Initial             &15.7                &7.4                 &20.9                &50.0                &23.5                &14.5 \hspace{0.15cm} &44.2                &29.9                &8.7                 &49.5                &48.8                &50.0                &41.2                &49.8                &50.0                &50.0                &7.7                 &13.1                &17.8                \\
Finetune            &\textbf{2.5}                 &\textbf{0.9}                 &\textbf{4.0}                 &\textbf{49.7}                &\textbf{6.6}                 &\textbf{6.9}  \hspace{0.15cm} &\textbf{37.1}                &\textbf{21.9}                &\textbf{5.0}                 &\textbf{49.2}                &\textbf{45.5}                &\textbf{49.9}                &\textbf{31.8}                &\textbf{47.3}                &\textbf{48.9}                &50.0                &\textbf{6.4}                 &12.8                &17.5                \\
REINFORCE           &11.7                &5.5                 &18.3                &50.0                &20.1                &11.5 \hspace{0.15cm} &43.3                &28.0                &8.0                 &49.5        &48.7                &50.0        &40.8                &49.5                &50.0                &50.0        &7.0                 &\textbf{12.6}                &\textbf{17.3}                \\
REI. Calib.    &15.3                &\worse{7.5}         &\worse{21.2}        &50.0                &22.9                &13.1 \hspace{0.15cm} &\worse{44.4}        &\worse{30.8}        &\worse{9.3}         &49.5        &\worse{49.1}        &50.0        &\worse{42.0}        &\worse{49.8}        &50.0                &50.0                &7.7         &12.9                &\worse{17.9}        \\
REI. t-DCF     &13.0                &6.0                 &18.9                &50.0                &21.2                &13.0 \hspace{0.15cm} &43.5                &28.3                &8.0                 &49.5        &48.7                &50.0        &40.6                &49.6                &50.0                &50.0        &7.3                 &12.8                &17.4                \\
REI. Calib. t-DCF&15.2                &7.1                 &20.6                &50.0                &23.0                &14.1 \hspace{0.15cm} &44.2                &29.8                &8.7         &49.5        &48.8        &50.0                &\worse{41.3}        &\worse{49.8}        &50.0                &50.0                &7.6                 &13.0                &17.7                \\
Soft t-DCF          &13.0                &6.0                 &18.9                &50.0                &21.2                &13.0 \hspace{0.15cm} &43.5                &28.2                &8.0                 &49.5        &48.6                &50.0        &40.5                &49.6                &50.0                &50.0                &7.2                 &12.8                &17.4                \\
\bottomrule
\end{tabular}
\end{table*}

\end{document}